# Long-term stabilization of carrier envelope phases of mid-infrared pulses for the precise detection of phase-sensitive responses to electromagnetic waves

T. Yamakawa[1,*], N. Sono[1,*], T, Kitao[1], T. Morimoto[1], N. Kida[1], T. Miyamoto[1,a], and H. Okamoto[1,2,b]

[1]Department of Advanced Materials Science, University of Tokyo, 5-1-5 Kashiwa-no-ha, Chiba 277-8561, Japan

[2]AIST-UTokyo Advanced Operando-Measurement Technology Open Innovation Laboratory (OPERANDO-OIL), National Institute of Advanced Industrial Science and Technology (AIST), 5-1-5 Kashiwa-no-ha, Chiba 277-8568, Japan

*equally contributed

[a]miyamoto@k.u-tokyo.ac.jp

[b]okamotoh@k.u-tokyo.ac.jp

Abstract

We report a newly designed mid-infrared-pump visible-probe measurement system, which can measure phase-sensitive responses to a mid-infrared pulse along the oscillating electromagnetic field. In this system, the pump light is a phase-locked mid-infrared pulse with temporal width of 100 fs, which is produced via difference frequency generation (DFG) from two idler pulses of two optical parametric amplifiers (OPAs) that are excited by the same Ti:sapphire regenerative amplifier. The probe pulse is a visible pulse with temporal width of 9 fs, and it is generated from a custom-built non-collinear OPA. By measuring the electric-field waveforms of mid-infrared pump pulses with electro-optic sampling and evaluating their carrier envelope phase (CEP) and the temporal positions of their envelopes relative to ultrashort visible probe pulses, we are able to perform double feedback corrections that eliminate both the following sources of drift. The CEP drift in mid-infrared pulses originating from fluctuations in the difference of optical-path lengths of the two idler pulses before the DFG is corrected by inserting a wedge plate in one idler path, and the drift in pump-probe delay times due to fluctuations in the difference of the overall optical-path lengths of the pump and probe pulses is corrected with mechanical delay lines. In this double-feedback system, the absolute carrier phase of mid-infrared pulses can be fixed within 200 mrad and errors in the measurement of phase-sensitive responses can be reduced to within 1 fs over a few tens of hours.

Recent developments in techniques for generating a strong terahertz (THz) pulse have opened up new ways using electromagnetic waves to control physical quantities like polarization and magnetization in solids[1,2]. The control of electronic phases using a THz pulse is also attractive, however only a few examples of using electromagnetic fields to induce phase transitions have been reported so far[3,4]. It is mainly because the amplitude of a THz pulse is generally difficult to enhance larger than 1 MV/cm, although several techniques have been proposed[5,6]. In a mid-infrared (MIR) region, it is easier to enhance electromagnetic-field amplitudes than in a THz region. Strong MIR pulses have indeed been used to induce a metal-superconductor transition in cuprates[7] and an insulator-metal transition in vanadium dioxide[8]. A strong MIR pulse can also induce novel nonlinear phenomena in solids. Studies have reported observations of non-perturbative responses in semiconductors[9-12], ultrafast polarization reversal in ferroelectrics[13], modulation of the on-site Coulomb interaction in Mott insulators[14], and sum-frequency excitation of Raman-active phonons in diamonds[15].

To reveal ultrafast electron and lattice dynamics in phase transitions and other nonlinear phenomena in solids with an MIR pulse, one must detect any changes to optical indexes induced by an oscillatory electric field. An MIR pump pulse with a stable carrier envelope phase (CEP) and a probe pulse with duration shorter than the period of the oscillatory fields are indispensable for monitoring these changes. CEP is defined by the phase of the carrier wave in a light pulse with respect to its envelope. In general, the CEP of the output of femtosecond laser sources such as a Ti:sapphire regenerative amplifier (RA) fluctuates unless a feedback device is introduced in the cavity[16]. To this end, several methods for passive stabilization of CEP during frequency down-conversion processes of the output of laser sources to an MIR pulse have been proposed: (1) inter-pulse differential frequency generation (DFG) between two frequency-detuned near-IR pulses generated separately by a common light source[17-19], (2) intra-pulse DFG of a spectrally broadened visible/near-IR pulse[20,21] or a two-color pulse generated in a dual-wavelength optical parametric amplifier (OPA)[22], and (3) four-wave-mixing with two-color filamentation[23].

Among those methods, the most general is the use of inter-pulse DFG using outputs from two OPAs[17], since it uses a relatively simple setup and offers wide tunability of the MIR-pulse frequencies. In this scheme, two signal (or idler) pulses with a constant phase difference $\Delta\phi$ are generated separately from two OPAs, in which both the excitation pulses and seed white-light pulses come from the same laser source. In an MIR pulse obtained via DFG processing of two signal (or idler) pulses, the original CEP fluctuations are cancelled. In an actual experimental setup, however, the optical-path-length difference between the two OPAs fluctuates due to jitter that originates from instabilities in temperature and air flow. This gives rise to fluctuations in the phase difference between the two OPA outputs, which result in the instability of CEP of the MIR pulse. For example, in the case that the photon energies of two OPA outputs are 0.7 eV and 0.6 eV, a change of 300 nm in the path-length difference (corresponding to a time difference of 1 fs) induces a change in the CEP of the MIR pulse (0.1 eV) of $0.3\pi$ rad or 7 fs in time. Note that the temporal change is magnified during the DFG process. The optical-path length of each OPA is typically 2 m, so that fluctuation of around $1.5 \times 10^{-7}$ of that length (= 300 nm/2 m) gives rise to a temporal error of 7 fs. Moreover, the path-length difference of an MIR pump pulse and a visible probe pulse in the pump-probe system will also fluctuate, leading to drift in the pump-probe delay time $t_\mathrm{d}$. The overall optical-path length in a pump-probe system is typically 8 m, so that the same degree of fluctuation in length $(1.5 \times 10^{-7})$ produces an error of 4 fs, which is comparable to the above-mentioned error due to fluctuations in the OPA path length. Therefore, active stabilization is indispensable for long-term experiments. For the stabilization of CEP in an MIR pulse, C. Manzoni *et al.* introduced a feedback correction device to an inter-pulse DFG system, in which phase stability of 110 mrad over 2 hours was achieved[24], although any fluctuations in the pump-probe delay were not eliminated. In case that measurements over a few tens of hours are necessary to detect CEP-sensitive phenomena, it is important to introduce feedback devices that stabilize both the CEP of the MIR pulse and the pump-probe delay.

In this letter, we report a newly designed feedback-control system for MIR-pump visible-probe measurements. This system measures the electric-field waveform of the MIR pump pulses with electro-optic sampling (EOS) and detects fluctuations in both CEP of the MIR pulse and the temporal position of its envelope peak relative to an ultrashort visible pulse. Hereafter, the latter is called the envelope peak position (EPP). Feedback devices are included that control the difference between the optical-path lengths of two OPAs and also the difference between the optical-path lengths of the pump and probe pulses. In the constructed system, error in the detection of a phase-sensitive response is reduced to as little as 1.0 fs over a few tens of hours.

Our setup is illustrated in Fig. 1(a). The light source is a Ti:sapphire RA, which generates a pulse with central wavelength (photon energy) of 800 nm (1.55 eV), duration of 35 fs, repetition rate of 1 kHz, and fluence of 7.5 mJ. The output from the RA is split into two. One is frequency-doubled in a $\beta$-BaB$_2$O$_4$ crystal and is used as the excitation pulse for a type-I non-collinear OPA (NOPA) with a $\beta$-BaB$_2$O$_4$ crystal[25]. In Fig. 1(b), we show the intensity profile of the NOPA output, obtained with a frequency-resolved optical-gate (FROG) using the retrieval algorithm[26]. The full width at half maximum (FWHM) of the pulse is 8.9 fs. Figure 1(c) shows the spectrum of the pulse, which ranges from 520 nm to 710 nm. The details of the NOPA were reported in ref. 25. The other output from the RA is introduced to a dual OPA that includes OPA1 and OPA2, in which a small part of the RA output is used to create a common white-light seed pulse and the residual is used to amplify the seed pulse in each OPA. By introducing two idler pulses from two OPAs to a 250-μm-thick GaSe crystal, an MIR pulse is generated via type-I DFG[27]. Although the CEP of the original RA output is unstable, the resulting MIR pulse is phase-stabilized[17]. In the experiments reported below, the wavelength and power of the idler pulse from OPA1 were fixed at 1640 nm and 390 μJ, respectively. The idler pulse from OPA2 was changed from 1900 nm to 2080 nm and its power was typically 120 μJ. Under these conditions, MIR pulses with the central frequency of 28-40 THz can be obtained.

The diameter of the MIR pulse thus obtained is expanded to 3 cm by two off-axis parabolic mirrors, OAP1 and OAP2 [Fig. 1(a)]. A beam splitter (BS) (a 500-μm-thick Si plate) is introduced to split the MIR pulse into two. The reflected MIR pulse with 60% the original intensity is used as the pump pulse for pump-probe measurements. Using OAP3, this pulse is focused to a spot 50 μm in diameter (FWHM), the position of which is hereafter referred to as the sample position. The transmitted MIR pulse is used in the EOS to measure its electric-field waveform. Using OAP4, the pulse is focused on a $LiGaS_2$ crystal, the position of which is hereafter referred to as the control position. The spot diameter at the control position is adjusted to be as large as that at the sample position. The visible probe pulse from the NOPA is also divided into two. Each pulse is focused through a hole drilled in the OAP onto the center of the sample or the control position with diameter of 20 μm. The former is used as the probe pulse in the pump-probe measurement and the latter as the sampling pulse for EOS.

A schematic of the EOS is shown in the lower-right part of Fig. 1(a). When an electric field is applied to a nonlinear optical crystal, the crystal's birefringence for the sampling pulse changes in proportion to the electric field through the Pockels effect, which can be measured as the difference in the signal of balanced photodiodes detecting the sampling pulse after passing through a quarter-wave plate and a polarizing BS. Thus, by changing the delay time of the sampling pulse relative to the MIR pulse, one can measure the electric-field waveform. We used a 20-μm-thick $LiGaS_2$ crystal, which can be used to detect electric fields with a wide range of frequencies. The electric-field waveform and Fourier power spectrum of a 33-THz MIR pulse are shown in Figs. 1(d) and (e), respectively. At this frequency, the maximum electric-field amplitude of 10.2 MV/cm is obtained. The electric-field amplitude is estimated from the pulse energy, the beam size, and the electric-field waveform [Fig. 1(d)] of the MIR pulse. The detection range is 3.3–12 μm (Supplementary Material S1). We performed consecutive EOSs to investigate the stability of the MIR pulses. The drift per hour was evaluated to be $0.5\pi$ (corresponding to 9 fs in the time domain) in CEP and 2 fs in the pump-probe delay.

The procedures for stabilizing the CEP and the pump-probe delay time $t_d$ are illustrated in Fig. 2(a). In the first step, the electric-field waveform of the MIR pulse at the control position is measured, which is then used as a reference and it is therefore called the reference waveform. In the second step, feedback is used to keep the electric-field waveform of the MIR pulse at the control position identical to the reference waveform in terms of both CEP and EPP. The stabilization of the EPP enables the stabilization of $t_d$. In the loop shown in step 2 in Fig. 2(a), the EOS and pump-probe measurements are performed simultaneously by changing the delay of the sampling pulse and that of the probe pulse to 300 fs, which takes about 30 seconds. This temporal range of measurements was chosen to strike a compromise between a short feedback interval and coverage of nearly the whole duration of the MIR pulse [see Fig. 1(d)].

To evaluate the drifts in CEP and EPP of an electric-field waveform at the control position relative to those of the reference waveform, we use the cross correlation of the two electric-field waveforms of the MIR pulses and its Hilbert transform. The drifts in CEP and EPP are illustrated in Fig. 2(b). CEP is expressed as the temporal difference between the carrier phase and the EPP multiplied by the carrier's angular frequency. The cross correlation of the reference waveform and the electric-field waveform in the $n$-th measurement (the $n$-th waveform) forms a fringe pattern. Examples of a reference waveform, an $n$-th waveform, and a cross-correlation profile measured with feedback enabled are shown in Figs. 3(a), (b), and (c), respectively. $\tau$ in the horizontal axis of Fig. 3(c) is the temporal parameter in the calculation of the cross-correlation profile and the phase at $\tau = 0$ gives the relative change in the carrier phase of the $n$-th waveform with respect to the phase of the reference waveform, i.e., the absolute carrier-phase drift. In Fig. 3(c), the absolute carrier-phase drift is almost equal to zero.

The EPP drift can be evaluated from the envelope of the cross-correlation profile of the reference and $n$-th waveforms. Assuming that the EPP varies slowly relative to the carrier wave, the envelope of the cross-correlation profile is equivalent to the cross-correlation profile of the envelopes of two waveforms (Supplemental Material S2). By extracting the envelopes of cross-correlation profiles using the Hilbert transform, we can determine the EPP of the $n$-th waveform. Calculating the cross-correlation profile before extracting the envelope tends to suppress noise, because uncorrelated noises are averaged out in the calculation. The evaluated EPP values are plotted with open circles in Fig. 3(d).

Because of the wide temporal width (FWHM~100 fs) of MIR pulses, the EPP data vary widely. Therefore, we next use a linear regression of EPP data from 300 consecutive cross-correlation profiles, which include information about waveforms during the last 2.5 hours of the experiment. The result is shown with the blue line in Fig. 3(d), in which variations in the raw data are suppressed and appropriate EPP values are obtained. This allows the precise determination of the drift of pump-probe delay. From the difference between the absolute carrier-phase drift and the EPP drift, we obtain the CEP drift [Fig. 2(b)].

Feedback for stabilizing both the CEP of the MIR pulse and the pump-probe delay time is accomplished as follows. The CEP is stabilized by inserting a pair of $CaF_2$ wedge plates (WP) after OPA2 [Fig. 1(a)]. The WP has a refractive index of 1.43 in the near-IR region and its wedge angle is 4°. One of the WPs is mounted on a motorized stage. By varying its insertion length, we can eliminate the path-length differences between the outputs of two OPAs due to fluctuations. The pump-probe delay time is stabilized by controlling two delay lines DL2 and DL3 [Fig. 1(a)] using EPP data.

To evaluate the performance of our system with the two kinds of feedback applied, we performed continuous EOS measurements at the sample position with and without feedback. The results are highlighted in Fig. 4(a) and (b), which show, respectively, the electric-field waveform of MIR pulses at the position shown by the broken line in (b) and those shown in a color contour when feedback is performed. In (b), all waveforms are normalized to the fluence of respective probe pulses. The result demonstrates that the phase is held stable for 20 hours. Figure 4(c) plots the absolute carrier-phase shifts with and without feedback for about 8 hours. Our feedback system works well, or the absolute carrier phase would never be stabilized. The stabilized absolute carrier phase has a root-mean-square error of 200 mrad over 20 hours, which corresponds to the drift of 1.0 fs in time. The symmetric distribution of the absolute phase jitter verifies that feedback calculated every 30 seconds is sufficient to reduce drift.

In summary, we constructed an MIR-pump visible-probe system in which double feedback correction is achieved. We evaluated the absolute carrier-phase drift as well as the drift of the envelope peak positions in MIR pulses by detecting electric-field waveforms. Feedback correction of both the carrier phase and the pump-probe delay allow us to hold the CEP constant within 200 mrad for at least 20 hours, which corresponds to a 1.0-fs error in time. Our highly stabilized system is a powerful tool for detecting phase-sensitive responses in solids, such as initial responses of electric-field-induced phase transitions, high-order nonlinear phenomena, and the formation of Floquet states by MIR pulses.

See Supplementary Material for the details of the electro-optical sampling and the determination of envelope peak position of MIR pulses using cross-correlation profile.

Acknowledgement

This work was supported in part by Grants-in-Aid for Scientific Research from the Japan Society for the Promotion of Science (JSPS) (Project Numbers: JP18H01166 and JP18K13476) and by CREST (Grant Number: JPMJCR1661), Japan Science and Technology Agency. T. Morimoto was supported by JSPS through the Program for Leading Graduate Schools (MERIT) and Research Fellowship for Young Scientists.

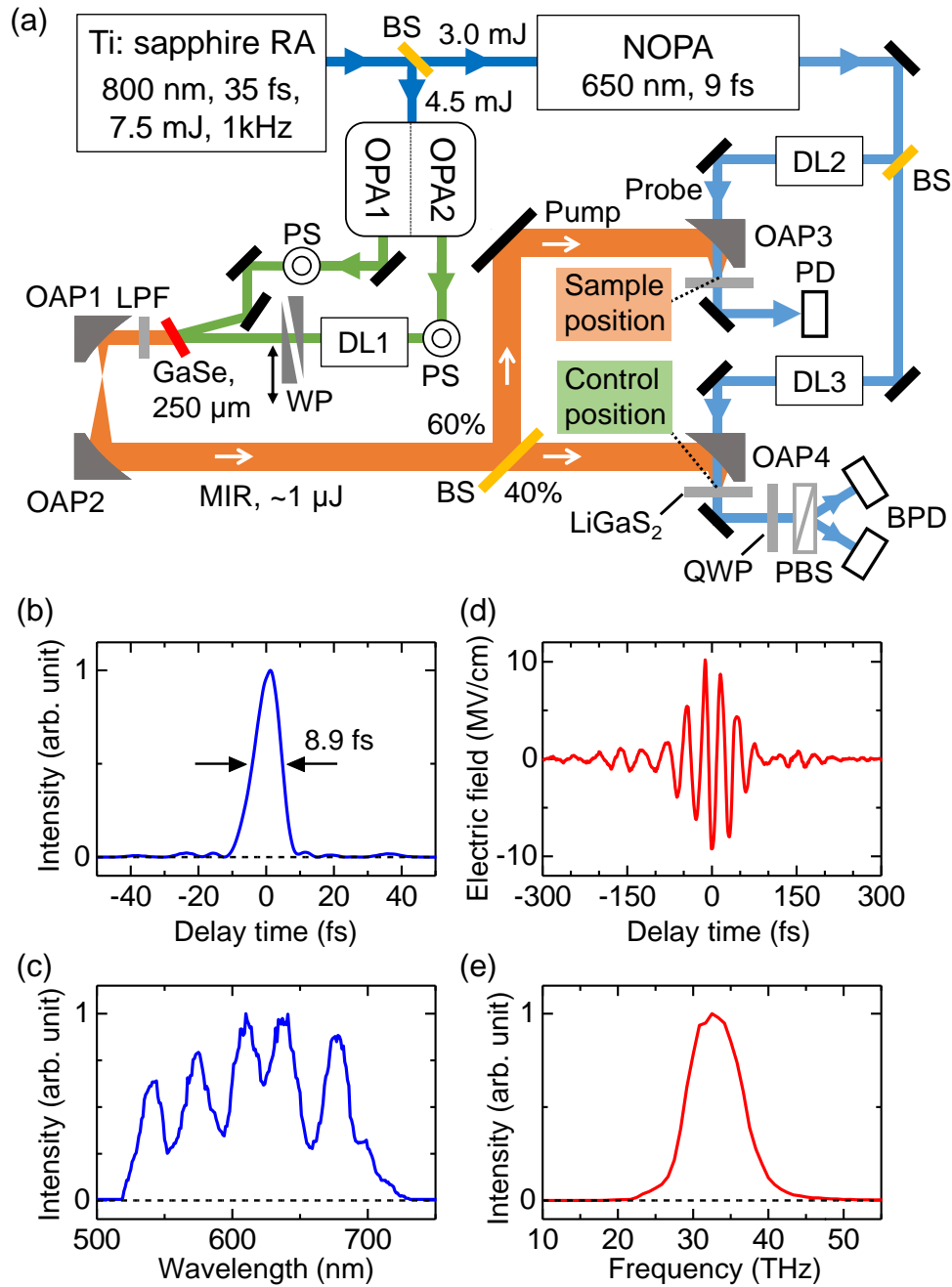

Fig. 1 (a) Setup of the MIR-pump visible-probe measurement system. OAP: off-axis parabolic mirror, BS: beam-splitter, PS: periscope, DL: delay line, WP: wedge-plate, PD: photodetector, QWP: quarter-wavelength plate, PBS: polarizing beam-splitter, BPD: balanced photodetector. (b) Typical intensity profile of an NOPA output retrieved from a FROG trace. (c) Spectrum of the NOPA output shown in (b). (d) The electric-field waveform of a 33-THz MIR pulse measured at the sample position. (e) Fourier power spectrum of the MIR pulse shown in (d).

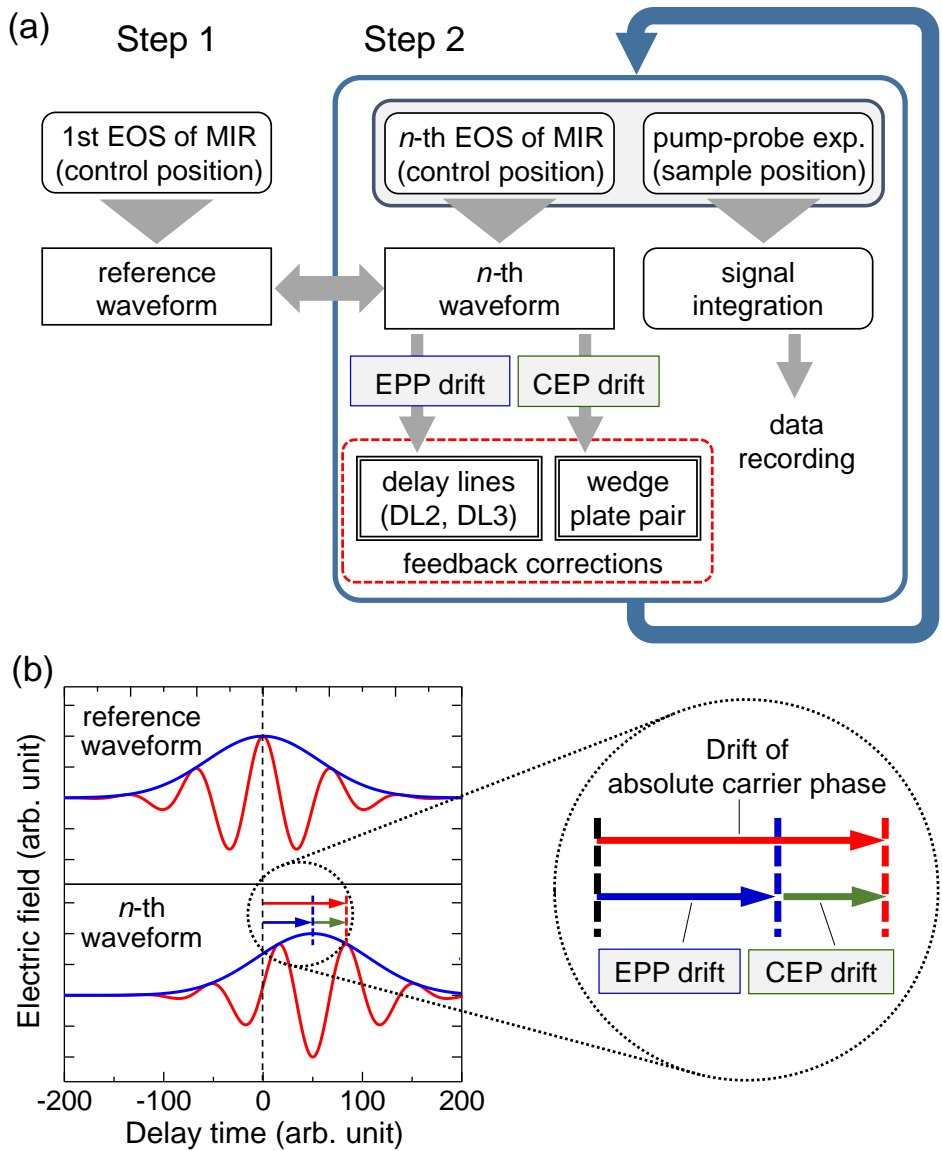

Fig. 2 (a) Diagram of double-feedback control (see text). (b) Schematic of drifts of absolute carrier phase, envelope peak position (EPP), and carrier envelope phase (CEP).

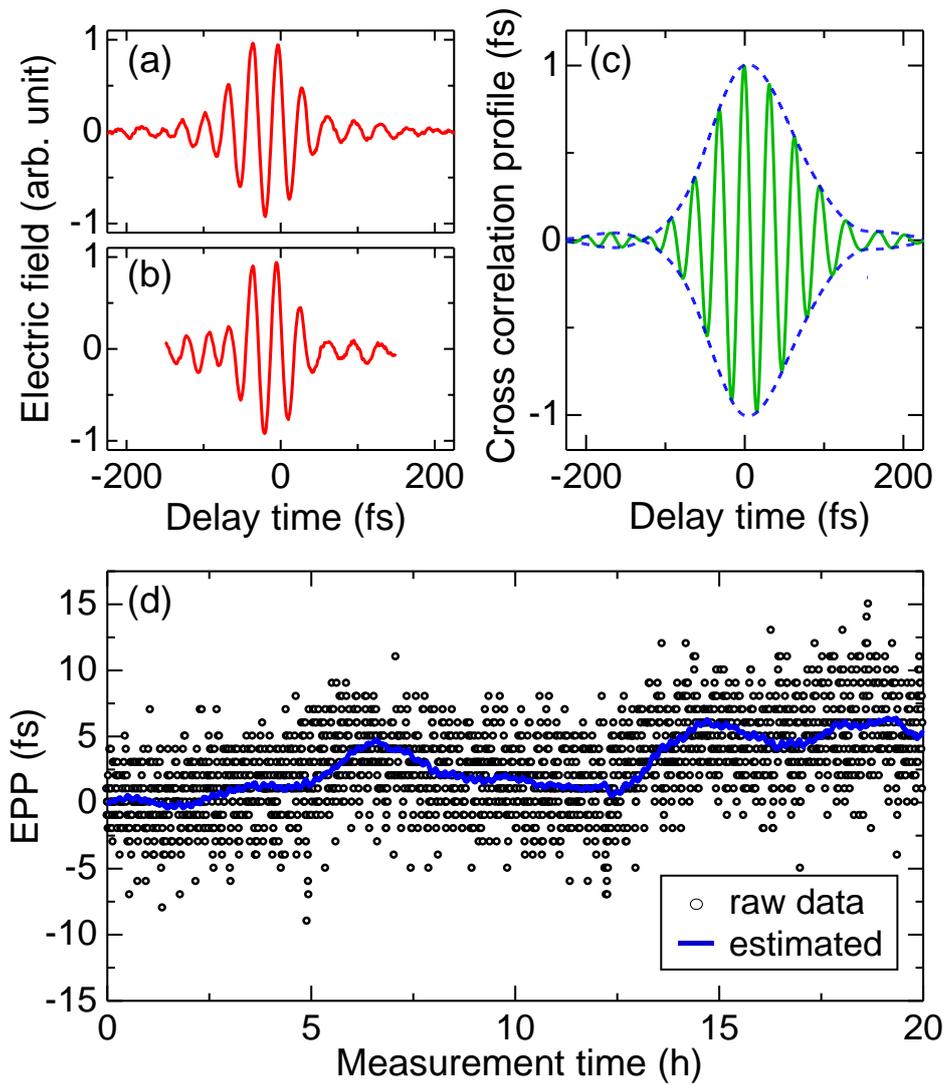

Fig. 3 (a,b) Electric-field waveforms of MIR pulses: (a) reference waveform and (b) *n*-th waveform. (c) Cross correlation profile between two electric-field waveforms shown in (a,b) and the envelope (broken line). (d) Envelope peak positions (EPPs) (black circles) and those estimated by the linear regression (blue line).

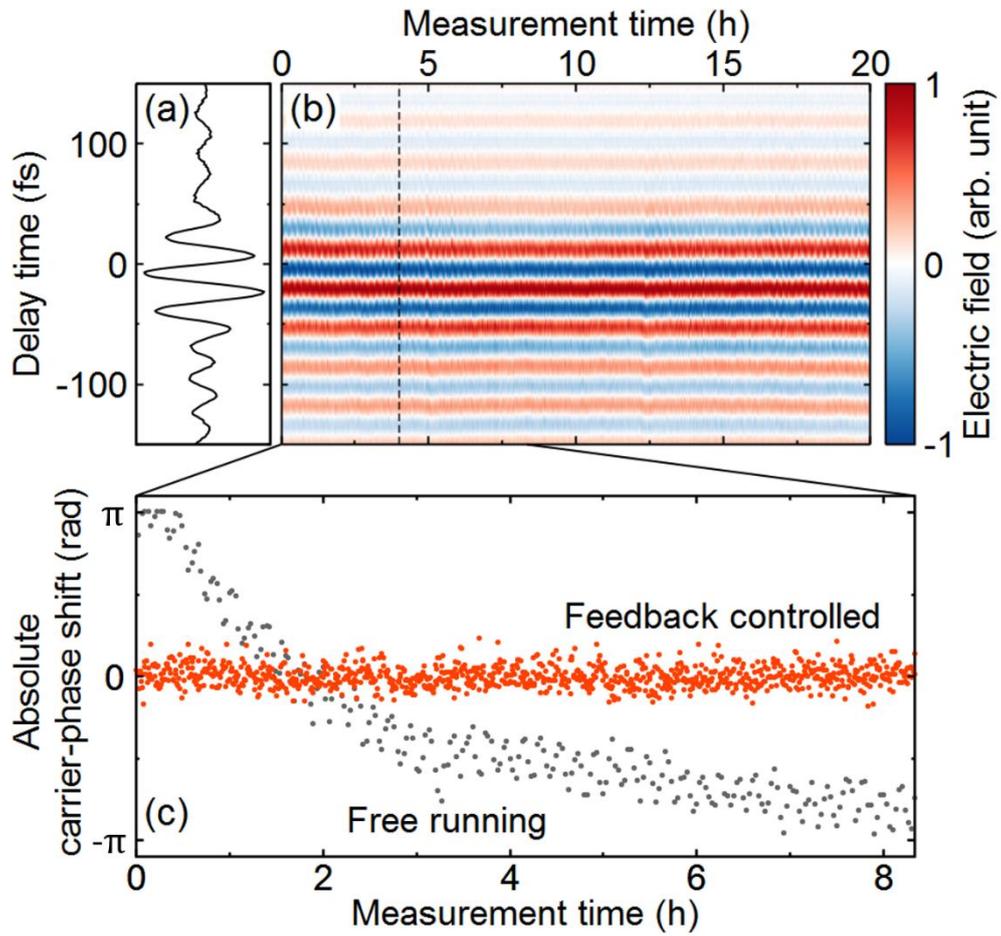

Fig. 4 (a) Electric-field waveform of MIR pulses measured at the temporal position marked with a broken line in (b). (b) Time evolution of normalized electric-field waveforms of MIR pulses. (c) Absolute carrier-phase shifts of MIR pulses with (orange circles) and without feedback (black circles).

**Supplemental Material**

**Long-term stabilization of carrier envelope phases of mid-infrared pulses for the precise detection of phase-sensitive responses to electromagnetic waves**

T. Yamakawa[1,*], N. Sono[1,*], T, Kitao[1], T. Morimoto[1], N. Kida[1], T. Miyamoto[1,a)] and H. Okamoto[1,2,b)]

[1]Department of Advanced Materials Science, University of Tokyo, 5-1-5 Kashiwa-no-ha, Chiba 277-8561, Japan

[2]AIST-UTokyo Advanced Operando-Measurement Technology Open Innovation Laboratory (OPERANDO-OIL), National Institute of Advanced Industrial Science and Technology (AIST), 5-1-5 Kashiwa-no-ha, Chiba 277-8568, Japan

*equally contributed

a)miyamoto@k.u-tokyo.ac.jp

b)okamotoh@k.u-tokyo.ac.jp

**Contents**

S1. Detectable frequency range of electro-optic sampling

S2. Determination of envelope peak position of MIR pulses using cross-correlation profile

## S1. Detectable frequency range of electro-optic sampling

In this section, we discuss the frequency range in which the electro-optic sampling (EOS) can be used. In EOS, the detection sensitivity of electric-field waveforms is dominated by the temporal width of a probe pulse, $\tau_p$. In the case that the period $T$ of an oscillatory electric field is much longer than $\tau_p$, the electric-field waveform is precisely measured. As $T$ approaches $\tau_p$, the sensitivity of the detection of the electric-field waveform decreases. However, by adding appropriate corrections of the detection sensitivity depending on $\tau_p/T$ to the data obtained by the EOS, we can evaluate a wide wavelength range of electric-field waveforms for $T \gtrsim \tau_p$. In our system in which the probe pulse with $\tau_p = 9$ fs is used, the shortest wavelength detectable is estimated to be 3.3 μm corresponding to $T = 11$ fs. This value is determined by the wavelength where the detection sensitivity is about 10% of that for a DC electric field. The longer wavelength bound is 12 μm, above which finite absorptions of a nonlinear optical crystal, LiGaS$_2$, exist. The other necessary condition for EOS is a phase matching between an MIR pulse and a sampling pulse. Since the LiGaS$_2$ crystal we used is very thin (20 μm thick), the phase-matching condition is fulfilled for 3.3 μm to 12 μm. In the region of 5.0-7.7 μm, the optics must be purged with dried air in order to avoid absorption by water vapor.

## S2. Determination of envelope peak position of MIR pulses using cross-correlation profile

A cross-correlation function $C_{fg}(\tau)$ between two time-dependent functions $f(t)$ and $g(t)$ can be calculated by

$$C_{fg}(\tau) = \int_{-\infty}^{\infty} f(t)g(t+\tau)\,dt. \tag{S.1}$$

We consider the case that $f(t)$ and $g(t)$ show electric-field pulses expressed as follows.

$$f(t) = A(t)\cos[\omega t + \varphi_1] \tag{S.2a}$$

$$g(t) = A(t - \tau_0)\cos[\omega(t - \tau_0) + \varphi_2] \tag{S.2b}$$

Here, $A(t)$ and $A(t - \tau_0)$ are the envelopes of $f(t)$ and $g(t)$, respectively, and $\tau_0$ is the envelope peak position (EPP) of $g(t)$ relative to that of $f(t)$. $\varphi_1$ and $\varphi_2$ are the carrier envelope phases (CEPs) of $f(t)$ and $g(t)$, respectively. Then, $C_{fg}(\tau)$ is expressed as

$$\begin{aligned}
C_{fg}(\tau) &= \int_{-\infty}^{\infty} A(t)A(t - \tau_0 + \tau)\cos[\omega t + \varphi_1]\cos[\omega(t - \tau_0 + \tau) + \varphi_2]\,dt \\
&= \int_{-\infty}^{\infty} A(t)A(t - \tau_0 + \tau) \\
&\quad \times \frac{\cos[\omega(-\tau_0 + \tau) - \varphi_1 + \varphi_2] - \cos[\omega(2t - \tau_0 + \tau) + \varphi_1 + \varphi_2]}{2}\,dt \\
&= \frac{1}{2}\cos[\omega(-\tau_0 + \tau) - \varphi_1 + \varphi_2]\int_{-\infty}^{\infty} A(t)A(t - \tau_0 + \tau)\,dt \\
&\quad + \frac{1}{2}\int_{-\infty}^{\infty} A(t)A(t - \tau_0 + \tau)\cos[\omega(2t - \tau_0 + \tau) + \varphi_1 + \varphi_2]\,dt.
\end{aligned} \tag{S.3}$$

The integration of the second term in the last line can be neglected when the envelope $A(t)$ varies slowly compared to an oscillation with frequency $2\omega$. Thereby, we obtain

$$C_{fg}(\tau) = \frac{1}{2}\cos[\omega(-\tau_0 + \tau) - \varphi_1 + \varphi_2]\int_{-\infty}^{\infty} A(t)A(t - \tau_0 + \tau)\,dt, \tag{S.4}$$

which is a product of an oscillation with the phase $(-\omega\tau_0 - \varphi_1 + \varphi_2)$ and the cross-correlation function of the envelopes of the two pulses. Here, the phase of the oscillation is equivalent to the carrier phase difference between $f(t)$ and $g(t)$. Next, we calculate the envelope of $C_{fg}(\tau)$ using the Hilbert transform, from which we can evaluate EPP ($\tau_0$) of $g(t)$ relative to that of $f(t)$. Thus, we obtain the CEP difference of the two waveforms, $(\varphi_1 - \varphi_2)$.

Now, let us return to equation (S.3). In the case of a Gaussian-shaped envelope:

$$A(t) = A \exp\left[-\frac{t^2}{2\sigma^2}\right], \tag{S.5}$$

the cross-correlation function in equation (S.3) has an analytical form of

$$\begin{aligned}C_{fg}(\tau) = &\frac{\sqrt{\pi}}{2} A^2 \sigma \exp\left[-\frac{(\tau - \tau_0)^2}{4\sigma^2}\right] \\ &\times [\cos(\omega(-\tau_0 + \tau) - \varphi_1 + \varphi_2) + \exp(-\omega^2\sigma^2)\cos(\varphi_1 + \varphi_2)].\end{aligned} \tag{S.6}$$

The contribution of the second term is very small as compared to the first term even for single cycle pulses ($\sigma\omega = 2\pi$), justifying the approximation of neglecting the second term in equation (S.3).